\documentclass[pra,showpacs,preprint,superscriptaddress]{revtex4-1}
\usepackage{amssymb}
\usepackage{graphicx}
\usepackage{bm}
\usepackage{epsfig}
\usepackage{amsfonts}
\usepackage{color}

\begin{document}

\title{Energy and lifetime of resonant states with real basis sets}

\author{Eloisa Cuestas}
\email{mecuestas@famaf.unc.edu.ar}
\affiliation{Facultad de Matem\'atica, Astronom\'{\i}a y F\'{\i}sica,
Universidad Nacional de C\'ordoba and IFEG-CONICET, Ciudad Universitaria,
X5016LAE C\'ordoba, Argentina}

\author{Pablo Serra}
\email{serra@famaf.unc.edu.ar}
\affiliation{Facultad de Matem\'atica, Astronom\'{\i}a y F\'{\i}sica,
Universidad Nacional de C\'ordoba and IFEG-CONICET, Ciudad Universitaria,
X5016LAE C\'ordoba, Argentina}

\affiliation{Qatar Environment and Energy Research Institute,
P.O. Box 5825, Doha, Qatar}

\begin{abstract}
Using a probabilistic interpretation of resonant states,
we propose a formula useful to calculate the lifetime of a resonance using 
 square-integrable real basis-set  expansion techniques. 
Our approach does not require an estimation of the density of states.
The method is illustrated with calculations of $s$ and $p$ resonant-state
energies and lifetimes.
\end{abstract}
\date{\today}

\maketitle

\section{Introduction}

Resonant states are of great importance in quantum physics since the pioneering 
work of Gamow modeling the $\alpha$ decay in radiative nuclei \cite{g28}. 
Resonant states appear in nuclear physics \cite{g28,be89}, 
atomic and molecular physics \cite{mc79,sc98}, and more
recently, in nanophysics \cite{bjs05,dbp08,fos09}. 

In a series of papers Hatano and collaborators  \cite{hatano08,hatano09,hatano10}
presented a probabilistic interpretation of  resonant states.
In these works, a resonance is defined as an eigenfunction of a Hamiltonian with
Siegert boundary conditions, that is, when the potential goes to zero, the wave 
functions is only an outgoing wave \cite{s39}.
Due to Siegert conditions, the problem is not Hermitian; the eigenvalues are
complex, the wave functions are not square integrable and
the particle number around a central volume exponentially 
decays in time because of momentum leaks from this volume. 

Many methods were developed in order to calculate the complex eigenvalues of
resonant states. The most widely used methods are based in complex scaling 
transformations that turn the resonance state into a square-integrable function
\cite{m98,moiseyev-book}. In particular, a  modification called
exterior complex scaling \cite{mbr04} was used recently in 
many photoionization problems \cite{prm09,tmr10}.

However,  due to its simplicity, there also exist several  techniques to calculate 
the real part  of a resonance eigenvalue using the Ritz-variational method with 
a square integrable real basis-set. 
These methods, called
in general stabilization methods \cite{mrt93}, use the abrupt change of a physical 
quantity when  a variable is ``crossing" a resonance. 
In particular different authors used the variational energies \cite{ht70,kh04}, 
the von Neumann entropy \cite{fos09}, or properties as the  double-orthogonality 
condition \cite{pots10} to obtain the real part of the energy.
In order to calculate the imaginary part of the resonant energy,  all these methods 
use an approximation of the density of states fitting the numerical
data using a Lorentzian function \cite{mrt93,kh04,pos10} .

In this paper we use the formalism of Hatano  \cite{hatano08,hatano09,hatano10}
 to obtain an expression of the
imaginary part of the energy useful to calculate it using only a stabilization method,
 without an extra fitting or approximation. 

The paper is organized as follows. In Section II we use the Hatano  formalism to
obtain a relation between the real and imaginary part of the energy and the density
of probability in the central region. 
In Section III we apply the results of the preceding section together with 
real basis-set expansions to calculate the complex energy of a resonance.
Finally, Section IV contains the conclusions with a
discussion of the most relevant points of our findings.

\section{Definition and properties of  resonances}

Here and elsewhere we use atomic units, $\hbar=1\,;m=1$. A general solution of the time-dependent Schr\"odinger 
equation 

\begin{equation}
\label{tdse}
H\psi(\vec{x},t)=\,i\,\frac{\partial}{\partial t} \psi(\vec{x},t)  \,,
\end{equation}

\noindent obeys a continuity equation 

\begin{equation}
\label{edc}
\frac{\partial\,\rho(\vec{x},t)}{\partial t}\,+\,\nabla \cdot
\vec{J}(\vec{x},t)\,=\,0 \,,
\end{equation}

\noindent where  $\rho$ and $J$ are the standard density and current of the 
probability \cite{merzbacher}

\begin{equation}
\label{udj}
\rho(\vec{x},t)\,=\,\left| \psi(\vec{x},t) \right|^2 \;\;;\;\; \vec{J}(\vec{x},t)\,=\,
\frac{1}{2 i} \left[ \psi^*(\vec{x},t) \nabla \psi(\vec{x},t)  -
\psi(\vec{x},t) \nabla \psi^*(\vec{x},t) \right] \,.
\end{equation}

\noindent Eq. (\ref{edc}) could be written  in integral form

\begin{equation}
\label{edci}
\frac{\partial}{\partial t}\,\int_\Omega \,\rho(\vec{x},t)\,d^dx\,=\,
-\,\int_{\partial \Omega} \, \vec{J}(\vec{x},t)  \cdot \vec{dS}\,,
\end{equation}

\noindent where  $d$ is the spatial dimension,
$\Omega$ an arbitrary volume, and $\partial \Omega$ its frontier.
Following Hatano \cite{hatano08,hatano09,hatano10}, we define $N_\Omega$, the number of particles  inside a 
volume $\Omega$ 

\begin{equation}
\label{npv}
N_\Omega(t)\,=\,\int_\Omega \,\rho(\vec{x},t)\,d^dx \,.
\end{equation}

\noindent From the definition of $J$ \cite{merzbacher}, Eq. (\ref{npv}) takes the form

\begin{equation}
\label{dndt}
\frac{\partial}{\partial t}\,N_\Omega(t)\,=\,-Re\left( \int_{\partial \Omega(t)}
\psi^*(\vec{x},t) \, \vec{p} \, \psi(\vec{x},t) \cdot d\vec{S} \right) \,.
\end{equation}

\noindent This equation, that express  the particle-number conservation inside the 
volume $\Omega$, corresponds to  Eq. (15) of reference \cite{hatano09}.
If the wave function goes to zero faster than $r^{-(d-1)}$ for large values of $r$ 
the RHS of Eq. (\ref{dndt}) goes to  zero, expressing the    conservation of the 
normalization, as happens for bound states.
Resonant eigenstates are not Hermitian solutions of the Schr\"odinger equation, and the wave functions have 
exponential divergences, Eq. (\ref{dndt}) describes a flux of particles outside any volume $\Omega$, even in the 
limit $\Omega \rightarrow \mathbb{R}^d$, and the  particle-number is not conserved.

Hatano suggested how to maintain the probabilistic interpretation of the wave function for resonant states 
\cite{hatano08,hatano09,hatano10}. The resonance  is interpreted as a metastable state, which is 
localized inside a volume $\Omega_0$ at $t=0$. Hatano defined a time dependent volume $\Omega(t)$ by the condition

\begin{equation}
\label{vt}
\frac{d}{d t}\,N_{\Omega(t)}(t)\,=\,0 \,.
\end{equation}

\noindent The reasonable initial condition for this equation is $N_{\Omega(0)}(0)\,=\,1$. This condition together  with
Eq. (\ref{vt}) imply that $N_{\Omega(t)}(t)\,=\,1\;\forall t\ge0$. The expectation value of an operator $\hat{{\cal O}}$ is calculated inside the volume $\Omega(t)$ as

\begin{equation}
\label{veo}
\langle \hat{{\cal O}}\rangle_{\Omega(t)}\,\equiv\,
\frac{\langle \psi \left|  \hat{{\cal O}} \right|\psi\rangle_{\Omega(t)}}{
\langle \psi \left. \right|\psi\rangle_{\Omega(t)}}\,,
\end{equation}

\noindent which is well defined for all times. 

Eqs. (\ref{npv}), (\ref{dndt}), and (\ref{vt}) give an equation for $\partial \Omega(t)$

\begin{equation}
\label{ef}
Re\left(\left\langle \psi \left| \left( \frac{\partial \vec{x}}{\partial t}\,-\,
\vec{p} \right) \right| \psi \right\rangle_{\partial \Omega(t)}  \right)\,=\,0\,.
\end{equation}

\noindent In order to  obtain solutions of  Eq. (\ref{ef}), we have to particularize the system. 
We restricted our study to   solutions of the time-dependent Schr\"odinger equation with fixed  energy 

\begin{equation}
\label{os}
\psi(\vec{x},t)\,=\,e^{-i E t }\,\psi_E(\vec{x}) \,,
\end{equation}

\noindent where $E$ and $\psi_E(\vec{x}) $ are the eigenvalue and eigenfunction of the
time-independent Schr\"odinger equation. 
We assume  one-particle Hamiltonians with  central potentials that
tend to zero at infinity.
Then, for central potentials we can use the reduced radial Schr\"odinger equation for 
$l$-waves

\begin{equation}
\label{rrse}
H_l u_{E,l}(r)=E\,u_{E,l}(r) \,,
\end{equation}

\noindent where

\begin{equation}
\label{rrh}
H_l \,=\,-\frac{1}{2}\,\frac{d^2}{d r^2}\,+\,\frac{l (l+1)}{2 r^2}\,+\,{\cal V}(r) \;,\;\;
\mbox{and}\;\;\;\psi_E(\vec{x})\,=\,\frac{u_{E,l}(r)}{r}\,Y_{l,m}(\Omega) \,,
\end{equation}

\noindent where $Y_{l,m}(\Omega)$ are the spherical harmonics \cite{merzbacher}.

Resonant states obey non-Hermitian  Siegert boundary conditions \cite{s39}, 
$u_{E,l}(r) \sim e^{i k r}$ for 
$r \rightarrow \infty$, where $k=\sqrt{2 E}$. Siegert states
are non normalizable, so they do not belong to the Hilbert space, and their complex 
eigenvalues $E\,=\, {\cal E}-i \Gamma/2 $ are interpreted as energies ${\cal E}$ 
and inverse lifetimes $\Gamma$ of metastable resonant states. 

By symmetry, the solution of Eq. (\ref{vt}) for the volume $\Omega(t)$ is a sphere
of radius $R(t)$,  $\Omega(t)=B(R(t))$, with the initial condition  $R_0=R(t=0)$. 
In this case, Eq. (\ref{ef}) gives the evolution of $R(t)$

\begin{equation}
\label{etrt}
\dot{R}(t)\,=\,\left. Im \left( \frac{\partial_r u_{E,l}(r)}{u_{E,l}(r)}\right) \right|_{r=R(t)}  \,.
\end{equation}

\noindent The RHS of this equation does not depend explicitly on $t$, and the formal solution is

\begin{equation}
\label{fst}
t\,=\,\int_{R(0)}^{R(t)}\, \frac{dr}{Im\left(\frac{\partial_r u_{E,l}(r)}{u_{E,l}(r)} \right) } \,.
\end{equation}

Even the resonant-states functions are not square-integrable, the initial condition
for $N_B$ fix the arbitrary normalization constant

\begin{equation}
\label{nc}
N_{B(R_0)}(0)\,=\,\int \left|\psi(\vec{x},0) \right|^2 d^3x\,=\,
\int_0^{R_0}\,\left| u_{E,l}(r)\right|^2 dr\,=\,1 \,.
\end{equation}

\noindent With this condition, the expression for $N_{B(R(t))}(t)$ takes the form

\begin{equation}
\label{ent}
N_{B(R(t))}(t)\,=\,e^{-\Gamma t}\,\int_{0}^{R(t)} dr \left| u_{E,l}(r) \right|^{2} \,=\,1\,,
\end{equation}

\noindent and the derivative with respect to $t$ gives

\begin{equation}
\label{gammagenerico}
\Gamma = \, \frac{ \left| u_{E,l} \left(R(t)\right)\right|^{2} }{\int_{0}^{R(t)} \;dr \left| u_{E,l} \left(r\right)\right|^{2}} \; \dot{R}(t) \;,
\end{equation}

\noindent combining this equation with Eq. (\ref{etrt}) we obtain

\begin{equation}
\label{gammagenerico2}
\Gamma = \, \frac{ Im \left(u^*_{E,l} \left(R(t)\right) \left. \partial_r u_{E,l}(r)\right|_{r=R(t)} \right)}
{\int_{0}^{R(t)} \;dr \left| u_{E,l} \left(r\right)\right|^{2}} \;.
\end{equation}

\noindent This simple expression for $\Gamma$ is not convenient for numerical calculation with real basis-sets.
With this purpose in view, we write the explicit expression for the  asymptotic 
behavior of the wave function

\begin{equation}
\label{ewfl}
 u_{E,l}(r)\,=\,C\,e^{i k r} \,v_l(k,r) \,,
\end{equation}

\noindent where $C$ is a normalization constant given by Eq. (\ref{nc}).  Eq. (\ref{ewfl}) in
Eq. (\ref{gammagenerico2}) gives 

\begin{equation}
\label{pi1}
\Gamma\, =\,\left(Im\;\left\lbrace i\,k + \left.\frac{\partial_{r}v_{l}(k,r)}{v_{l}(k,r)}\right| _{r=R(t)}\right\rbrace \right)\,
\frac{\left| u_{E,l} (R(t)) \right|^{2}}{\int_{0}^{R(t)} \; dr \left| u_{E,l} (r) \right|^{2}} \,.
\end{equation}

\noindent By  definition $k^2= 2 E = 2 {\cal E}- i \Gamma$, which gives 

\begin{equation}
\label{rek}
Im(i \,k)\,=\,Re(k)\,=\,\left[\sqrt{{\cal E}^2+ \left(\frac{\Gamma}{2}\right)^2}+{\cal E} 
\right]^{1/2}\,.
\end{equation}

\noindent  Finally, using Eq. (\ref{rek}) in Eq. (\ref{pi1}),  we arrive at the expression

\begin{equation}
\label{fgamma}
\Gamma\, =\,\left( 
\left[ {\cal E} + \sqrt{ {\cal E}^{2}+\left( \frac{\Gamma}{2}\right)^{2} }\right]^{\frac{1}{2}} + 
\mbox{Im} \left.\left( \frac{ \frac{d}{dr}v_l(k,r)}{v_l(k,r)} \right) \right|_{r=R(t)} \right)
\frac{\left| u_{E,l} (R(t)) \right|^{2}}{\int_{0}^{R(t)} \;dr \left| u_{E,l} \left(r\right)\right|^{2}}   \;.
\end{equation}

For the important case treated in references \cite{hatano08,hatano09} of  potentials  with a  finite support, 

\begin{equation}
\label{potsc}
{\cal V}(r)\,=\,\left\{ \begin{array}{ll} V(r) & \mbox{if}\,r < r_0   \\
0 &\mbox{if}\,r > r_0  \end{array}  \right. \,,
\end{equation}

\noindent we have  an explicit expression of $v_l$ for $l$-waves, valid for $r>r_0$

\begin{equation}
\label{vl}
v_l(k,r) \,=\,\sum_{j=0}^l \,(-1)^j \frac{(l+j)!}{j! (l-j)!}\,
\frac{1}{(2 i k r)^j}\,.
\end{equation}

\noindent  In particular, for  $s$-waves, $l=0$, $v_0(k,r)=1$, and then
Eq. (\ref{fgamma})  reduces to a linear relation   for $\Gamma^2$, which gives

\begin{equation}
\label{cpik}
\Gamma\,=\,\left[2\, {\cal E} \,+\,\left(\frac{\left|  u_{E,0} (R(t)) \right|^2}{2\,
\int_{0}^{R(t)} \; dr \left| u_{E,0} (r) \right|^{2}}
\right)^2 \right]^{1/2}\,\frac{\left|  u_{E,0} (R(t)) \right|^2}{\int_{0}^{R(t)} \; dr \left| u_{E,0} (r) \right|^{2}} \,.
\end{equation}

\noindent Equation (\ref{fgamma}), and for $l=0$, Eq. (\ref{cpik}) relate $\Gamma$ with real magnitudes, 
$ {\cal E}$ and $\left|  u_{E,l} (R) \right|^2$. This fact makes these
equations useful tools to calculate $\Gamma$ using square integrable real basis sets, as we show in the next section.

\section{The Ritz-variational method and resonances: Numerical Expansions}
\label{sec-var}
	
In this section we use Eqs. (\ref{fgamma}) and (\ref{cpik}) to calculate the imaginary part of the energy
of a resonant state,  applying the
Ritz-variational method with real square-integrable basis sets.

We exploit three facts of the variational expansion: 

{\bf ($i$)} There are several accurate
methods to calculate the real part of the eigenvalue, or energy,  ${\cal E}$, of a resonant  state.

\begin{figure}
\begin{center}
\includegraphics[width=15.5cm]{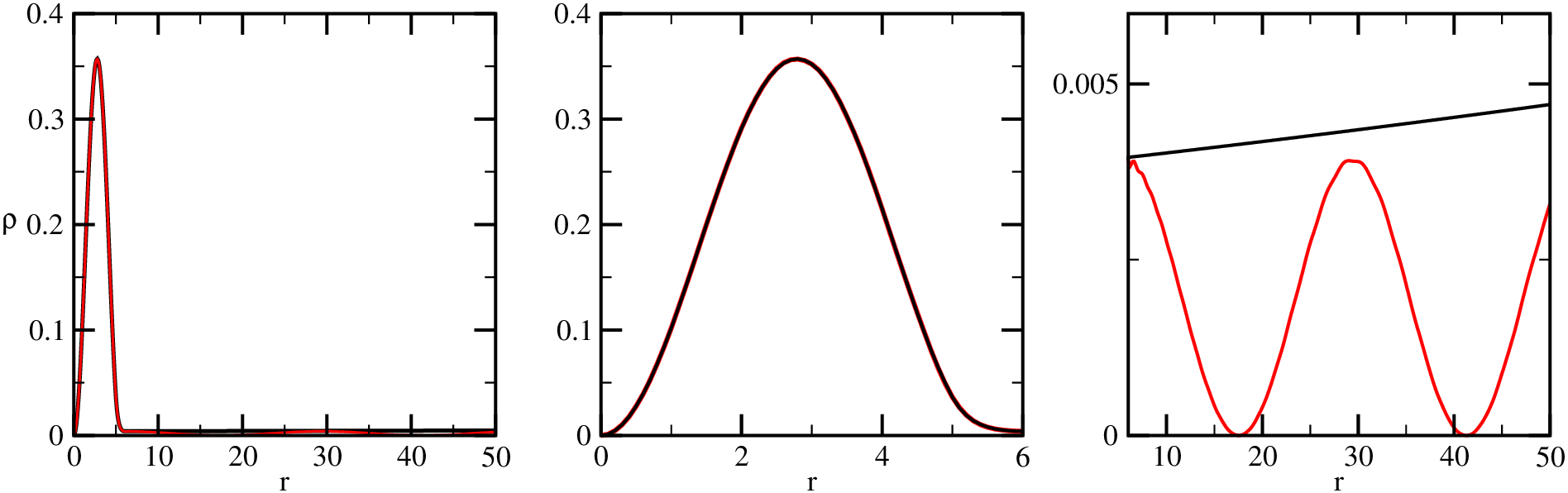}
\end{center}
\caption{  \label{rho} (color online) $\rho(r)$ for a $l=0$-resonant state, exact 
(black line) and  Ritz-DO approximation
with $N=100$ (red line) of the potential from Eq. (\ref{wpb}).  (a) global view. (b) Region $0\leq r \leq r_0=6$, where
the resonant state is localized,  and (c) The exterior region $r \ge r_0=6$, where the resonance diverges and
the variational expansion is a stationary wave. }
\end{figure}

{\bf ($ii$)} The variational method gives good approximations to the exact
(non-normalizable)  densities $\rho(r)$ where the resonant states are localized
(see figure \ref{rho}).

{\bf ($iii$)}   Eq.  (\ref{fgamma}) involves
just real quantities that could be evaluated with Ritz-variational wave functions.

Even Eqs. (\ref{fgamma}) and  (\ref{cpik}) are valid for $R\ge r_0$, because item ($ii$),
 in the numerical calculations we take $R=R_0=r_0$, and then 
$\int_0^{R_0} \; dr \left| u_{E,l} (r) \right|^{2} =1$.

\subsection{$l=0$}

We begin with the simple case of  $s-$waves, $l=0$

For a clear notation, we will omit the subindexes $E$ and $l=0$, then
 $u(r)$ for an arbitrary potential  $V(r)$ has the form

\begin{equation}
\label{psiei}
u(r)\,=\,\left\{ \begin{array}{ll} u^<(r) & \mbox{if}\,r < r_0   \\
u^<(r_0)\,e^{i k (r-r_0)}
 &\mbox{if}\,r > r_0  \end{array}  \right. \,.
\end{equation}

\noindent Then, in this case  Eq. (\ref{cpik}) takes the form

\begin{equation}
\label{gsw}
\Gamma\,=\,\left[2\, {\cal E} \,+\,\left(\frac{\left| u^<(r_0) \right|^2}{2}
\right)^2 \right]^{1/2}\,\left| u^<(r_0) \right|^2 \,.
\end{equation}

\noindent Eq. (\ref{gsw}) involves two real quantities, $\left| u^<(r_0) \right|^2$ and
$ {\cal E}$, both quantities are well approximated by applying the Ritz method using a real 
square-integrable basis set truncated at order $N$, $\{\Phi_i\}_1^N$. 
In this approximation, the Hamiltonian $H_l$ is replaced by a $N \times N$ Hermitian 
matrix $[H_l]_{i,j}=\langle \Phi_i| H_l |\Phi_j \rangle$, and
we obtain $N$ eigenvalues $E_n$ and eigenvectors $\vec{a}^{(n)}$. The corresponding
orthonormal eigenfunctions are

\begin{equation}
\label{xi}
\psi_n^{(N)}(r)\,=\,\sum_{i=1}^N\,a^{(n)}_i \,\Phi_i(r) \;\;\;;\;\;\;
\sum_{i=1}^N\,\left(a^{(n)}_i\right)^2\,=\,1 \;;\;
n=1,\ldots,N \,.
\end{equation}

\begin{figure}
\begin{center}
\includegraphics[width=15.5cm]{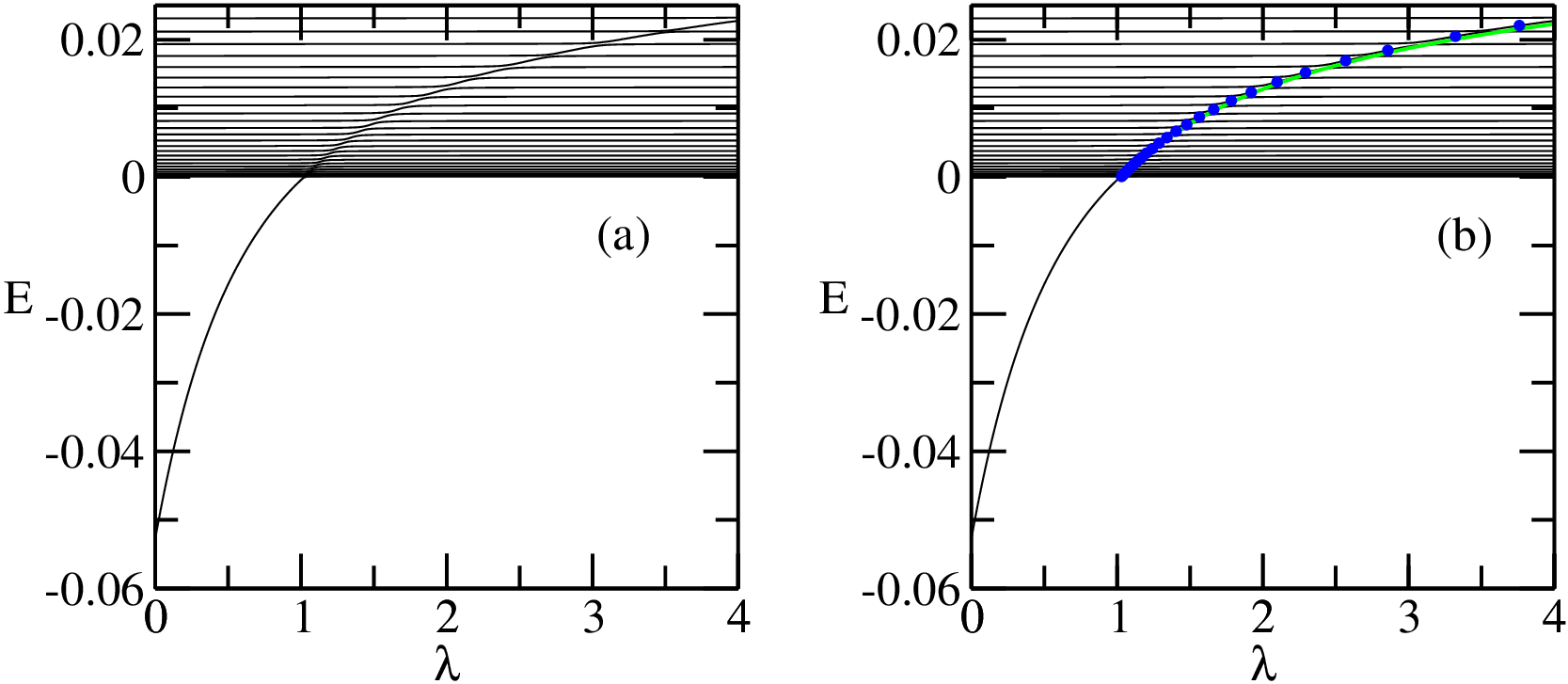}
\end{center}
\caption{  \label{epbl0} (color online) (a) The first 30 eigenvalues of the $N=100$ 
Hamiltonian matrix
of the $l=0$ block of the potential Eq. (\ref{wpb}) as a function of the barrier height  $\lambda$. 
(b) Same as (a) plus exact (green line) and approximate (blue dots) energies of the resonant state.}
\end{figure}

\noindent  In particular, we used the double-orthogonality method (DO) \cite{pots10} 
in order to calculate the real part of the resonant energy ${\cal E}^{(N)}$ and the 
variational square-integrable approximation to
the resonant wave function $u^{(N)}(r)$. This method assumes that the potential
depends on a parameter $\lambda$, and when $\lambda$ is varied on an interval
$[\lambda_L,\lambda_R]$ a given eigenvalue $n_0$ crosses the resonant energy value
at $\lambda_{n_0}$, as illustrated in figure \ref{epbl0}. The method uses the fact that in both sides of the
interval, $\lambda_L$ and $\lambda_R$, the $n_0$ eigenvalues and eigenvectors 
in the left and in the right of the avoiding-crossing zone
correspond to different states of the quasi-continuum and the resonant state is 
orthogonal  to both of them. We define the double-orthogonality function

\begin{equation}
\label{dort}
D_n(\lambda) = |\langle \psi_n(\lambda_{L}), \psi_n(\lambda)   \rangle|^2 +
|\langle \psi_n(\lambda_{R}), \psi_n(\lambda)\rangle|^2,\;\;\; \mbox{for} \;
\lambda_{L} < \lambda < \lambda_{R} \,.
\end{equation}

\begin{figure}
\begin{center}
\includegraphics[width=10.5cm]{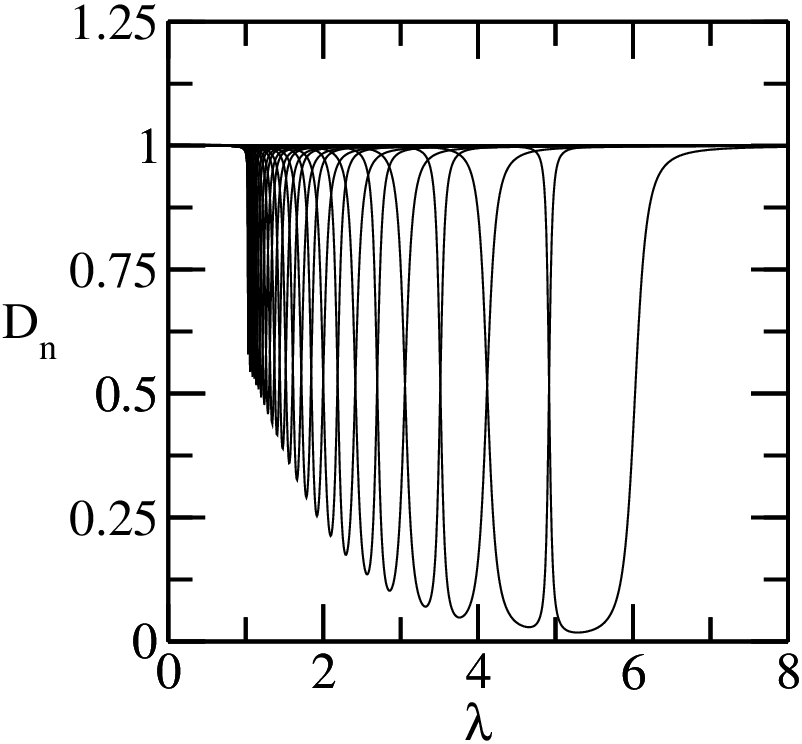}
\end{center}
\caption{  \label{do} $D_n$ for the  $l=0$ block of the potential
Eq. (\ref{wpb}) as a function of the barrier height $\lambda$
for $N=100$ and  $n=2,\ldots, 30 $. The minimum of each curve is defined as the 
localization of the resonant state.}
\end{figure}

\noindent Because the eigenfunctions are normalized, $0 \leq D_n(\lambda) \leq 2$.
For a given eigenvalue $n_0$, we define the localization of the resonance 
$\lambda_{n_0}$ as the value of 
$\lambda$ where $D_{n_0}(\lambda)$ reaches its 
minimum, that is, where the eigenfunction has a minimum projection onto to the  
quasi-continuum states, at it is shown in figures \ref{do} and \ref{dol1}. 
This method has the advantage over other stabilization methods in
that we have to solve the variational problem just a single time. 
The price we pay is that
$\lambda_n$ is also an output of the method and we cannot choose it arbitrarily.
The best approximation of the resonance is defined as 

\begin{equation}
\label{edo}
\lambda_{n_0}\,=\,\min_{\lambda \epsilon [\lambda_L,\lambda_R]} \,D_{n_0}(\lambda)
\;\;\;;\;\;\;{\cal E}(\lambda_{n_0})\,=\, E_{n_0}(\lambda_{n_0})
\end{equation}

Once we determine the optimal $\lambda_{n_0}$, its eigenfunction 
$\psi_{n_0}^{(N)}(r)$ \cite{pots10} together  the normalization condition 
Eq. (\ref{nc}) give

\begin{equation}
\label{ar}
u^{(N)}(r)\,=\,\frac{\psi_{n_0}^{(N)}(r)}{\int_0^{r_0}\,\left|\psi^{(N)}_{n_0}(r)\right|^2\,dr}\,=\,
\frac{\sum_{i=1}^N\,a^{(N)}_i \,\Phi_i(r)}{\sqrt{\sum_{i,j=1}^N\,a^{(N)}_i a^{(N)}_j  I_{i,j}(r_0)}}\,,
\end{equation}

\noindent where
\begin{equation}
\label{imn}
 I_{i,j}(R)\,=\,\int_0^{R}\,\Phi_m(r)\,\Phi_n(r)\,dr\,.
\end{equation}

\noindent  Then, from Eq. (\ref{gsw}) we obtain for $\Gamma^{(N)} $

\begin{equation}
\label{gl0}
\Gamma^{(N)} \,=\,\frac{\left( \psi_{n_0}^{(N)}(r_0)\right)^2}{
\sum_{i,j=1}^N\,a^{(N)}_i a^{(N)}_j  I_{i,j}(r_0)}   \; \left[ \left(
\frac{\left( \psi_{n_0}^{(N)}(r_0)\right)^2}{2\,
\sum_{i,j=1}^N\,a^{(N)}_i a^{(N)}_j  I_{i,j}(r_0)} \right)^2\,+\, 2\, {\cal E} 
 \right]^{1/2}\;.
\end{equation}

As a particular case, we  calculated  resonant states for an exactly solved 
problem,  the well+barrier  potential \cite{hogreve95}

\begin{equation}
\label{wpb}
{\cal V}(r)\,=\,\left\{ \begin{array}{ll} -V_0 & \mbox{if}\,r < \Delta   \\
+\lambda  &\mbox{if}\, \Delta<r<r_0  \end{array}  \right. \,,
\end{equation}

\noindent where all the parameters are positive.  The exact wave functions are different
combinations of exponential  functions in each sector,
with continuous logarithm derivative at $r=\Delta$ and $r=r_0$. The exact energies for bound, virtual and
resonant states are given as solutions of three different  transcendental algebraic equations, which are
obtained applying  the corresponding boundary condition at $r=r_0$  \cite{hogreve95}.

\begin{figure}
\begin{center}
\includegraphics[width=10.5cm]{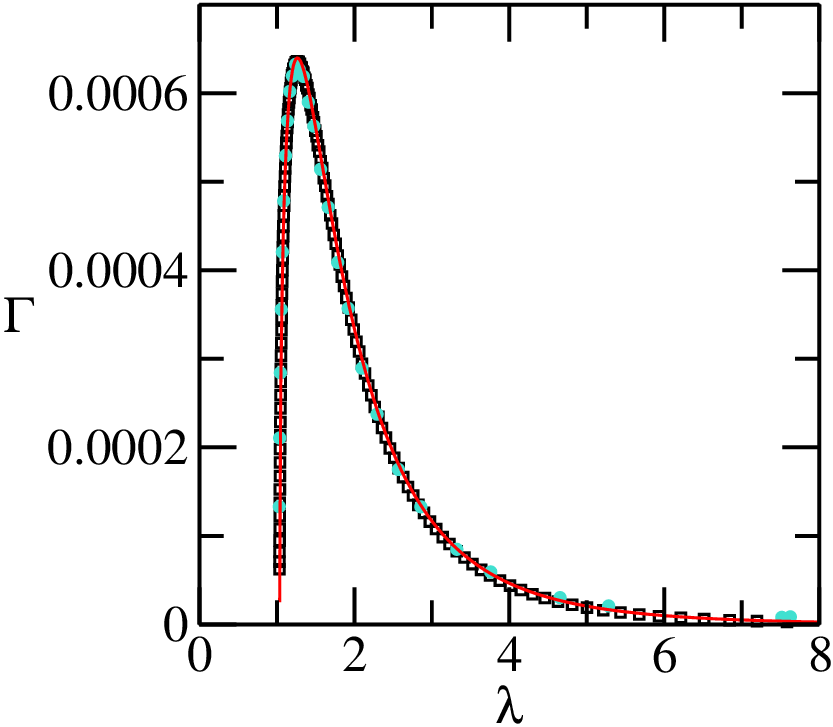}
\end{center}
\caption{  \label{gammapbl0}  (color online) Exact (red line), $N=100$ (turquoise dots)  and $N=500$ 
(black squares) values of $\Gamma$ against the barrier height  $\lambda$ for  the $l=0$ block of the potential 
Eq. (\ref{wpb}).}
\end{figure}

The calculations were done as a function of the barrier height  $\lambda$,      
with fixed values of $V_0=0.15,\,\Delta=5$, and $r_0=6$. 
A convenient orthonormal basis set is given by

\begin{equation}
\label{lbs}
\Phi_i(r)\,=\,\frac{1}{\sqrt{(i+1) (i+2)}}\,e^{-r/2} L^{(2)}_i(r)  \;;\; i=1,\ldots,N \,. 
\end{equation}

\noindent where $L^{(2)}_i(r)$ is the Laguerre polynomial of degree $i$ and order 2
\cite{abramowitz}. 

 In figure \ref{epbl0}(a) we show the first thirty eigenvalues given by the Ritz
method  with $N=100$. In figure \ref{epbl0}(b) we add the exact resonant energy curve and the 
values calculated with the $DO_n$ functions, which are shown in figure \ref{do}
for $N=100$, and $n=2,\ldots,30$. Once the energy 
 was obtained, the width $\Gamma$ of the resonance is calculated using Eq. (\ref{gl0}).
In figure  \ref{gammapbl0} we show $\Gamma(\lambda_n)$ for two different sizes of the basis set:
$N=100$, for $n=2,\ldots,30$, and $N=500$, for $n=2,\ldots, 140$. Our data show an excellent agreement
with the exact curve  $\Gamma(\lambda)$, that  is also included  in the figure.

\subsection{$l=1$}

From  Eq. (\ref{vl}), the function  $v_l(k,r)$ for $p$-waves takes the form

\begin{equation}
\label{vl1}
v_1(k,r_0)\,=\,1+\frac{i}{k r_0} \,.
\end{equation}

In this case, it is convenient to re-write Eq. (\ref{fgamma}) as a function of
$x\equiv Im(k)$ 

\begin{equation}
\label{fgammarw}
x^3 \,+\,\left(\frac{1}{r_0} \,+\,\frac{\left|u(r_0)\right|^2}{2}\right)\,x^2\,+\,
\left({\cal E} \,+\,\frac{1}{2 r_0^2} \,+\,\frac{\left|u(r_0)\right|^2}{2 r_0}\right)\,x
\,+\,{\cal E}\, \frac{\left|u(r_0)\right|^2}{2} \,=\,0 \,.
\end{equation}

\noindent The definition of resonance establish that $Im(k)>0$, 
and we  proved for each case that we studied 
that Eq. (\ref{fgammarw}) has an unique positive root.
Finally  $\Gamma$ is obtained from the definition of $k$ as

\begin{equation}
\label{gk}
\Gamma\,=\,-2\,x_p\,\sqrt{2\,{\cal E} \,+\,x_p^2} \,,
\end{equation}

\noindent where $x_p$ is the positive root of Eq. (\ref{fgammarw}). 

\begin{figure}
\begin{center}
\includegraphics[width=10.5cm]{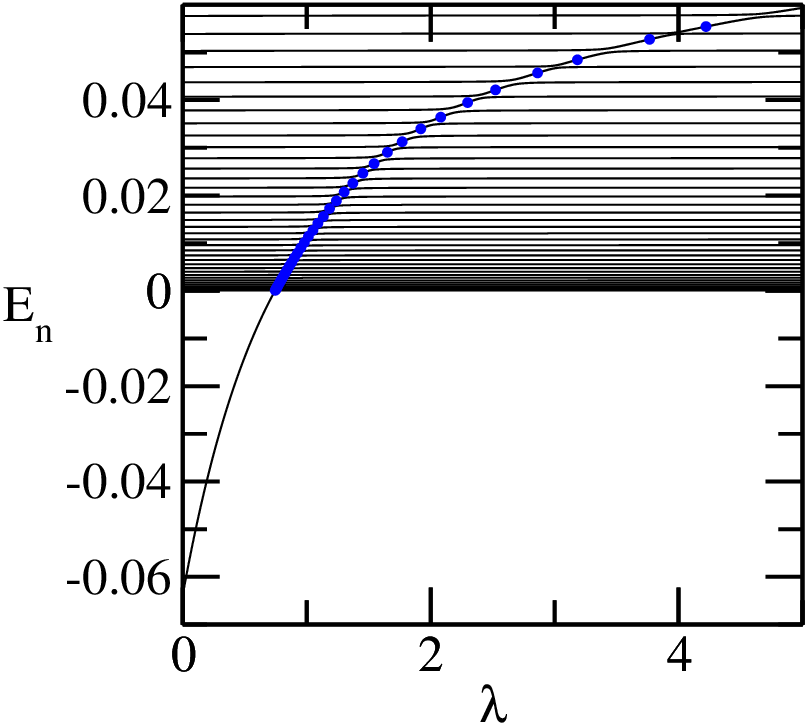}
\end{center}
\caption{  \label{epbl1} (color online) The first 30 eigenvalues of the $N=100$
Hamiltonian matrix
of the $l=1$ block of the potential Eq. (\ref{wpb}) as a function of the barrier height  $\lambda$.
The approximate  energies of the resonant state calculated with DO are also
shown (blue dots).
 }
\end{figure}

\begin{figure}
\begin{center}
\includegraphics[width=10.5cm]{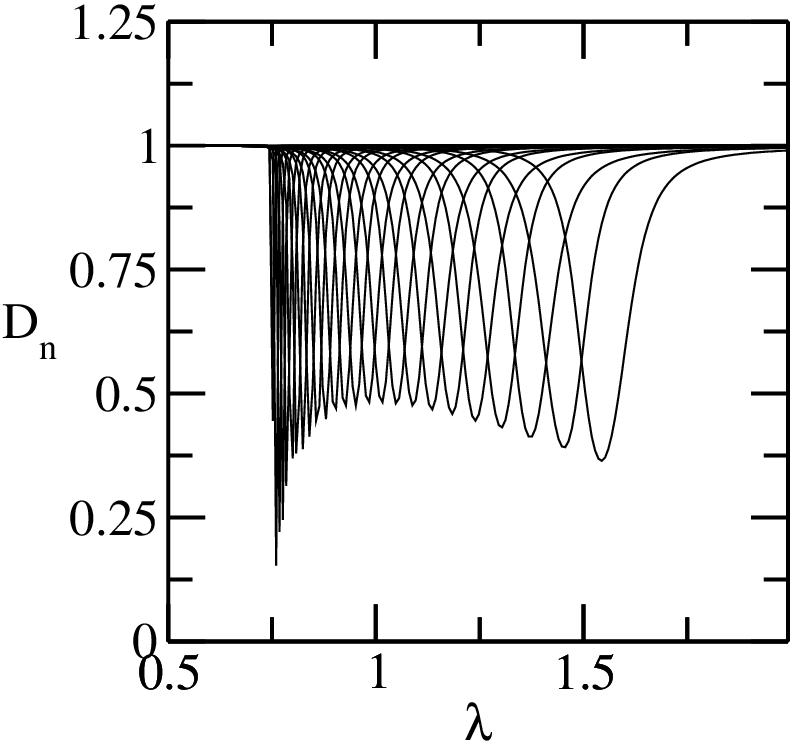}
\end{center}
\caption{  \label{dol1} $D_n$ for the  $l=1$ block of the potential
Eq. (\ref{wpb}) as a function of the barrier height $\lambda$
for $N=100$ and  $n=2,\ldots, 30 $. The minimum of each curve is defined as the
localization of the resonant state.}
\end{figure}

\begin{figure}
\begin{center}
\includegraphics[width=10.5cm]{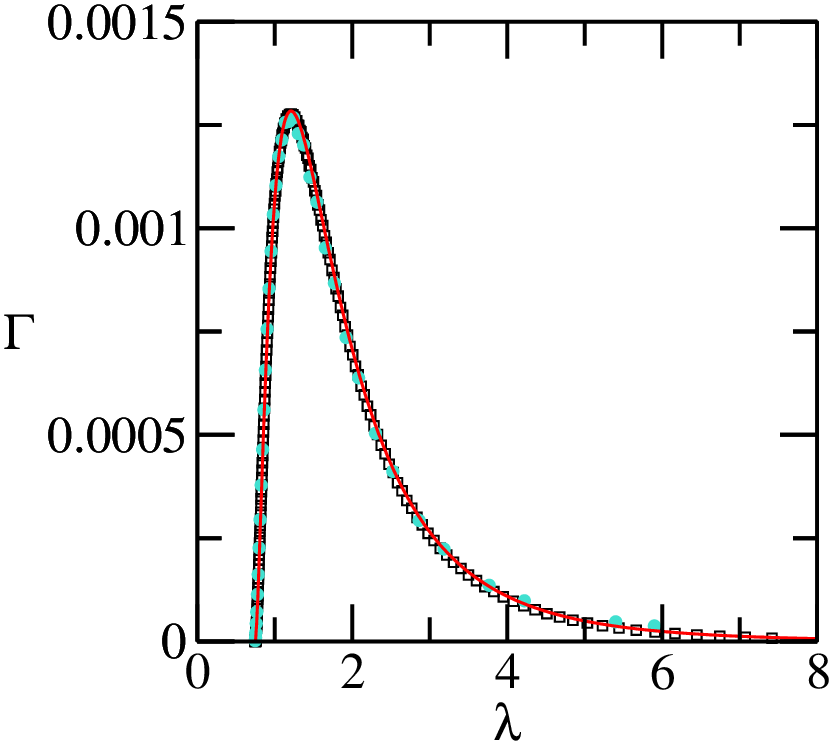}
\end{center}
\caption{  \label{gammapbl1}  (color online) 
 Exact (red line), $N=100$ (turquoise dots)  and $N=500$
(black squares) values of $\Gamma$ against the barrier height  $\lambda$ for  the $l=1$ block of the potential Eq. (\ref{wpb}).}
\end{figure}

We calculated the inverse lifetime $\Gamma$ for a $l=1$ resonant state of the potential
Eq. (\ref{wpb}) as a function of the barrier height  $\lambda$, 
with fixed values of $V_0=0.3,\,\Delta=5$, and $r_0=6$. 
In figure \ref{epbl1} we show the first thirty eigenvalues of the 
$N=100$ $p$-block of the Hamiltonian matrix and the resonant energies ${\cal E}$
calculated with the DO method.  The curves $D_n(\lambda)$ for $n=2,\ldots,30$ are shown 
in figure \ref{dol1}. 
Note the qualitative differences between the DO curves for $l=0$ and $l=1$ 
in figures \ref{do} and \ref{dol1} respectively. 
These differences are
due to the existence of a virtual state between the bound and the resonant states for 
the $l=0$ case, which it is absent in the $l=1$ case, where the bound state is continued 
directly in  a resonant state.

In figure \ref{gammapbl1} we
show the exact curve $\Gamma$ vs. $\lambda$ and the approximate values obtained with
 two different
basis-set size:  $N=100$, for $n=2,\ldots,40$ and $N=500$, for $n=2,\ldots,200$.
As in the case with zero angular momentum, we
obtain an excellent agreement between exact and approximate results.

\section{conclusions}

In this work we used a probabilistic interpretation of resonant (Siegert) states 
based on the conservation of the number of particles inside a time-dependent volume
\cite{hatano08,hatano09,hatano10}. 
The advantage of the probabilistic interpretation of resonant states is that 
is possible to work with Siegert states in a similar way 
to bound states, calculating probabilities, expectation values, etc.

In particular, we obtain the exact equation (\ref{fgamma}), which reduces to 
Eq. (\ref{cpik}) and Eq. (\ref{fgammarw}) for $l=0$ and $l=1$ respectively.
These equations relate the inverse lifetime 
with other real magnitudes, the energy and the density of a resonance.
In previous papers, once  the energy of a resonance is obtained applying a
real algebra stabilization-like method, the resonance width is calculated performing a fitting of the 
density of states (\cite{kh04} and references therein, \cite{pots10}). 
In the present work Eq. (\ref{fgamma}) gives a value for  $\Gamma$ with the
same degree of accuracy that we obtain for the energy of the resonance ${\cal E}$.

We emphasize the simplicity of the calculations compared to other methods that use 
complex algebra to study resonant states.
We present our results for potentials with finite support, but Eq. (\ref{fgamma}) is 
valid  in  general, and the
calculation of $\Gamma$ could be  corrected by a systematic perturbative expansion.

An open question is if Eq. (\ref{fgamma}) could be generalized for few-particle systems, where
many channels are present. We are working in this direction.

\acknowledgments
We would like to thank Jacob  Biamonte for a critical reading of the manuscript. We acknowledge  SECYT-UNC,  CONICET and MinCyT C\'ordoba
for partial financial support of this project.

\end{document}